# On the Verge of One Petabyte - the Story Behind the BaBar Database System

Adeyemi Adesanya, Tofigh Azemoon, Jacek Becla, Andrew Hanushevsky, Adil Hasan, Wilko Kroeger, Artem Trunov, Daniel Wang
*SLAC, Stanford, CA 94085, USA*

Igor Gaponenko, Simon Patton, David Quarrie
*LBNL, Berkley, CA 94720, USA*

On the behalf of BaBar Computing Group

The BaBar database has pioneered the use of a commercial ODBMS within the HEP community. The unique object-oriented architecture of Objectivity/DB has made it possible to manage over 700 terabytes of production data generated since May'99, making the BaBar database the world's largest known database. The ongoing development includes new features, addressing the ever-increasing luminosity of the detector as well as other changing physics requirements. Significant efforts are focused on reducing space requirements and operational costs. The paper discusses our experience with developing a large scale database system, emphasizing universal aspects which may be applied to any large scale system, independently of underlying technology used.

## 1. INTRODUCTION

The BaBar experiment at SLAC has been in production since May 1999. The design of the BaBar Database has been detailed at many conferences, including the past CHEP conferences ([5], [7]). This paper covers our experience with supporting persistency for the BaBar experiment: a multi-hundred terabyte database system. It discusses achievements, issues, and new development, with main focus on period since the last CHEP in September 2001.

## 2. PROVIDING PERSISTENCY FOR BABAR

BaBar is one of the largest HEP experiments currently in production. The system is growing very fast, both in size and complexity. The amount of production data has doubled in the past year and half. Now it exceeds 700 terabytes of unique production data, making it the largest known database in the world and drawing media attention.

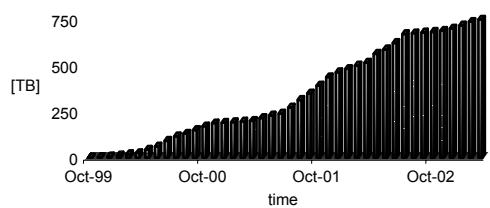

Figure 1. BaBar data growth

The system is used at SLAC, the location of the BaBar experiment, as well as over 25 collaborating sites, including newly established Tier A site in Padova (INFN, Italy), which is responsible for event reprocessing. The growing distribution of the system allows for more efficient use of resources spread across many sites in ten countries.

The database system glues the BaBar software - millions of lines of complex C++ code — together with powerful hardware. This hardware includes many tens of data servers, tens of terabytes of disk space, hundreds of terabytes of tertiary storage, over a hundred secondary servers[1] and thousands of CPUs, where user jobs are executed. This unique position of the database system requires it to be as robust and scalable as possible. On the other hand, the growth in complexity, demands, and size forces the system to constantly expand in every direction:

- Distribution (number of sites, data exchange rates),
- Address space (number of servers, disks, files),
- Data transfers (disk I/O, LAN and WAN), and
- Number of clients (perhaps the most important).

The number of clients is particularly important, since it scales with the number of connections to servers, the number of open files, and locks, disk traffic, network traffic and so on.

Continuous system expansion inevitably means exploring new limits and hitting unforeseen bottlenecks on regular basis. Many of the hard issues are ultimately found to be core-database independent.

The lively environment, continuous expansion and changing requirements are far from expected 'stable-production' mode of a mature, 3-year old experiment many were initially expecting.

## 3. CHANGES AND CHALLENGES

### 3.1. Prompt Reconstruction

Prompt Reconstruction (PR) has continued to be the most critical component of the BaBar data processing system, requiring the most attention and effort from the BaBar Database group in the past year. However, the focus has started to shift away from PR in the last few months, proving that PR is now designed and optimized to scale without requiring constant attention from the database experts.

---

[1] "*Secondary*" servers include lock servers, journal servers, and catalog servers

The next sub-chapters describe in details issues related to PR. The issues are arranged in chronological order.

### 3.1.1. Number of physics streams

Originally, the output from PR was persisted into four output physics streams. In September 2000, a decision was made to readjust event selection, and replace four streams with twenty. While this change improves event selection and makes data analysis much easier, it introduces an extra burden on the database system. Each stream is associated with a set of database files[2]. Data server performance degrades with number of open files and TCP connections. Moving from four to twenty streams introduces five times the load on data servers and effectively means five times the opened files, and five times the connections to data servers. Similarly, the number of streams scales with the number of locks, and lock server performance also decreases with increasing number of locks.

Regaining performance and scalability has required serious work including code changes, tuning servers and clients, as well as adding more hardware.

### 3.1.2. Introducing skim collections

In mid-2002 a decision was made to redo event selection, and produce only four output streams instead of twenty. In addition to the four streams, PR started to produce 115 "skim" collections. A skim collection is a pointer collection. Introducing 115 skims allowed physicists to further improve event selection, but once again, the database system limits were stretches, and the system was faced with a challenge how to support 20 times more collections. Pre-creating many thousands of collections (120 per processing node) started to take long time and reduced efficiency. Also, a central tree node hierarchy addressing all collections in a given federation started to reach physical limits (64K pages per container), since the system was not designed to scale to hundreds of millions of collections of one type. Significant changes to the code had to be made to overcome these limits. Luckily, encountering the limit was an artifact of our design, even though the 64K page limit is imposed by the Objectivity system. Thus we were in control of the code that imposed it. Skim collections also necessitated revisiting collection metadata. Having to maintain backwards compatibility with the existing system significantly increased the complexity of the redesign.

### 3.1.3. Introducing Linux clients

To reduce cost of BaBar computing, and to improve performance of PR farms by taking advantage of emerging fast x86 CPUs, Linux farms were introduced alongside existing, slower, Solaris *Netra T1s*. While the gain in speed was impressive, SLAC made an unlucky hardware choice. Frequent hardware failures -- unexpected random reboots -- visibly reduced the efficiency of PR production farms, forcing us to look for new ways to manage recovery from crashes and the resulting orphan locks associated with unusable machines.

One issue was exposed by frequent client crashes: many connections were left open on the server side after the crashes. This was due to the fact that Objectivity servers are not using TCP option "KEEPALIVE". Frequent client crashes cause rapid accumulation of obsolete connections, and degraded server performance, forcing us to restart servers more frequently.

### 3.1.4. Restructuring

Restructuring PR farms was one of the most important steps towards removing the scalability problems we were constantly fighting in the previous years. The original design allowed for only one run to be processed at any given time. In order to keep up with increasing incoming data rates, we were constantly forced to increase the number of clients and servers. Scaling this system to a couple of hundred nodes was successful, but daunting. After that number, adding more clients became inefficient, and the failure rate dramatically increased.

The solution was to restructure PR: separate rolling calibration phase from reconstruction phase, and allow processing multiple runs to be processed in parallel. Introducing these changes allowed the system to scale far beyond rates expected in the future. The new system is now scalable; however, it is also an order of magnitude more complex structurally. This requires a new approach to manage it. See [3] for further details.

### 3.1.5. Turning off raw and rec

The experiment decided to stop persisting raw and rec components. Keeping data in these formats was found to be very inefficient. Because of their bulky size (50kB for raw and 150kB for rec), the cost of maintaining them was high (disk i/o, tapes) given how infrequently they were used. Another factor against keeping raw and rec was introduction of the "new mini" ([2]), which manages to virtually replace most needed functionality of raw and rec, while keeping the size in 10kB range.

Turning off raw and rec was one of the very few decisions that *reduced* stress on the database system.

### 3.1.6. Setting up farms in Padova

To offload SLAC resources, BaBar collaborators from INFN in Italy offered help in running event reprocessing production. The facility was setup in Padova in early 2002, consisting of several large 100% Linux-based processing farms. This was the very first attempt to run PR production outside of SLAC.

Introducing Linux servers required us to port the developed in-house extensions to the data server (AMS) to the new platform. Numerous platform dependent optimizations had to be made to achieve satisfactory performance.

Due to a substantially different setup in Padova, we faced many problems that initially appeared to be database related, but in fact were purely related to the local infrastructure, to a particularly different NFS setup. Since Padova took the lead and started to stress the latest BaBar releases before we were able to do so at SLAC, new release-related problems were often first uncovered there, slowing down the process of setting up their PR farms.

---

[2] 8 different types: col, evt, evshdr, tag, aod, esd, raw, rec

The Padova PR production farms have been in stable production since August 2002, and database-related troubleshooting effort is now greatly reduced.

### 3.2. Analysis

#### 3.2.1. Bridge technology

Deploying bridge technology ([6]) was one of the largest changes in the BaBar analysis world since last CHEP. Bridge technology enabled us to address multiple "*slave*" federations, by providing a mapping (bridge) between collection names and slave federations that contain these collections. In other words, it allowed us to bind all of the data together, and make it appear to the client as if it were in just one federation. Before bridge federations were deployed, many users learned the hard way how difficult it was to track, which collections are in which federations.

The latest bridge technology includes deep copy – extraction of collections into a separate federation. This feature was initially expected to be widely used by the export tools, in particular BdbServer++ ([4]). It has not yet been used extensively.

To maintain logical separation of data, separate bridge federations were set up for each BaBar Run. Also, simulated and real data are kept separately. Thus, to cover 3 Runs (BaBar is in the middle of Run 3 at the moment), we needed six federations.

#### 3.2.2. Expansion

The analysis farm was significantly expanded. It now consists of over a hundred slave federations spread across 66 terabytes of disk space attached to 29 data servers, and supported by 34 secondary servers.

#### 3.2.3. Reducing lock collisions

Independently of skim collection production by PR farms, data is skimmed again in the analysis federations, and new skim collections are produced there directly by hundreds of production skim jobs running 24x7. These jobs constantly need update locks for central collection metadata resources, often colliding with user jobs accessing collection metadata in read mode (while locating a collection). These problems were known and present in the system for some time. Recently they were exacerbated by increased number of skim jobs running simultaneously. Fortunately, tuning the code and reducing access time to precious metadata resources dramatically alleviated the problem.

#### 3.2.4. Restarting servers and outages

Data availability has been significantly improved since first couple of years in production. New, non-disruptive operational procedures have been developed. Data is spread across many federations, so we can take offline only one (or few) federations at a time. This significantly improved data availability. Now, data is available over 96% of time on average, and the largest contributors to outage time are power outages.

The problem related to KEEPALIVE described in the PR section had even bigger impact in the analysis world than it had in PR. To balance load, data from many federations is usually spread across many servers, and one server handles data for many federations. Thus restarting an AMS in order to clean up obsolete connections often requires a disruptive outage affecting many federations.

A similar problem was also found in the lock server, which had a hard-coded limit of connections (1K). Hitting this limit was very disruptive, especially for a bridge federation's lock server. We have worked with Objectivity on solving that problem. The limit was raised, and the KEEPALIVE flag was enabled.

### 3.3. SP

Simulation production (SP) grew visibly from 8 sites in 2001 to 25. Reorganizing SP production (combining three stages into one) allowed us to reduce amount of transferred data about ten times. This balanced the increases related to:
- increased luminosity, and
- doubled number of generated Monte Carlo events per every real event.

### 3.4. Data corruption

In the past year we identified a relatively large number of corrupted database files. The first problem was a race condition in the code that closes and reopens a file. This was identified and fixed in the Objectivity kernel. This issue was related to ~10% of corruption cases.

The majority of cases happened in the second half of the year 2002. The corruption unexpectedly almost stopped occurring after a major power outage at SLAC in mid December. At approximately the same time we identified and removed several Linux client machines from production that were "misbehaving", i.e. the number of lost packets was very high for these machines (>0.12%) even under a light load. Non-BaBar, non-database code was used to demonstrate this behavior. Unfortunately we still do not understand the real cause of these corruptions, and what "fixed" them. We are working hard on understanding and eliminating the remaining issues that affect a few percent of the data.

The third source of corruption was relatively well contained, as it happened only in the Prompt Calibration farm. It was related to incorrect updates of B-Tree indices. This problem was understood by our experts, and promptly fixed by Objectivity in the next version of their product.

## 4. NEW FEATURES

### 4.1. Redesigned Condition Database

In the past few years we have discovered many deficiencies of the existing Condition Database (CDB) design – its performance started to degrade, and it lacked many important features.

#### 4.1.1. Issues with "old" CDB

In the old CDB, performance of both update and lookup operations with metadata (at the level of "interval containers") was exponentially deteriorating with the number of "versions" and total number of "intervals/objects" in a container. For example, more rolling calibrations stored in the corresponding containers resulted in slower response of the CDB. As a result the "finalize" time of PR jobs got longer and longer. This was known as "staircase" problem. The origin of the general metadata inefficiency was found to be caused by inefficient implementation of multi-dimension indices in Objectivity. We temporarily solved that by manually "purging" the metadata, which usually cost us long outages at the PR farms, ranging from few hours up to one day.

Due to the monolithic design of the metadata we had to manually "merge" (synchronize) conditions data produced by Prompt Reconstruction and several Reprocessing farms. The direct result was that we could not increase the number of Condition/DB installations in the production area to cope with the increased luminosity.

We also had to introduce the OID Server. Its role was to reduce the metadata traffic between clients and AMS-es serving condition databases.

In practice, the old CDB did not allow us to have more than about 100k objects per condition. Supporting even that meant slow access time and long modification time.

In order to address these problems, we embarked on a redesign.

#### 4.1.2. The "new" CDB

The main features of the new, redesigned CDB system consist of:
- new model of metadata: 2-d space of validity and insertion time, revisions, persistent configurations, types of conditions and hierarchical namespace for conditions
- state ID
- flexible user data clustering
- support for distributed update and use
- improved scalability
- significant performance improvements (x100-1000 speedup for critical use cases)
- improved robustness.

The new system was deployed in September 2002. Existing CDB data was converted to the new format, and the "masterCDB" federation was established. The performance has greatly improved since then, and maintenance of the system has become dramatically easier. For instance, the inter-federations sweeps are now very quick – in the order of a few minutes, compared to many hours. We manage to maintain 16 CDB production federations, while before, managing 4 was an extremely difficult tasks.

After a few months of gaining experience and managing the new setup, we have started to develop a set of management tools that would further simplify and automate CDB management.

Independently, an effort is made to reduce size of the CDB.

#### 4.1.3. Reducing CDB size

The full condition snapshot currently occupies ~40GB, so it seems like nothing compared to hundreds of terabytes of event store data. However, there is a growing need to support simple, laptop-analysis without any dependency on the central services. This means users should be able to import CDB locally to their laptops. Since the size of current commodity disks is still on the order of few tens of gigabytes, 40GB is prohibitively big. We have started to make several steps to address this problem.

One way of reducing size is through compression. Compressing the full snapshot reduces its size down to ~11GB. More details on compression may be found in the following chapter.

Another way to support laptop-analysis is to recluster condition objects. Re-clustering could help in two ways. We could:
- Extract a subset of conditions for specific type of analysis, for instance micro-analysis. This has already been demonstrated: a snapshot for micro-analysis occupies 200MB. A simple, but powerful active-schema based tool has been developed and used to achieve that.
- Recluster conditions, and get rid of old CDB data, which is stored purely for backward compatibility with old releases.. We are expecting to recover about 10-15GB of CDB by doing that.

### 4.2. Reducing disk cost

The growth of size of BaBar data sample continues to be fast. Although persisting raw and rec – two largest components of event – has been turned off, there are still significant other contributors to that growth:
- Luminosity increases
- Doubled number of SP events per real event
- Reprocessed data: each year all data taken so far since October'99 is reprocessed

Since disk cost is the most expensive part of the BaBar computing, a large focus is put on reducing disk space usage in order to reduce the already-high costs of BaBar computing. Several measures were made to address that.

#### 4.2.1. Compression

A standard Objectivity installation does not support data compression. We worked with Objectivity on the 'hooks'

that would allow us to plug in compression into their system. Next we developed in-house a plug-in compression library.

Decompression is easily configurable and can happen either on the server side, or the client side. To not overload servers, we opted for client side decompression.

In tests, our production data compression ratio reaches 2:1, which means significant reduction in disk usage. We also showed that introducing compression would actually slightly speed up job execution. In our environment, most jobs are I/O bound, and random disk reads are a dominating factor. Introducing compression means an extra step that now needs to be done: de-compression, but it also means that much less data need to be read from very busy disks.

Unfortunately, after making detailed analysis of access patterns, we realized that most of the on-disk data is not referenced often enough to justify paying the cost of compressing it, and it is better to use dynamic staging, and purge off disk files that have not been used for few days.

Because of this realization, we are still not compressing production data. We are in the process of estimating whether we could compress all micro-data and fit it all on disk in a compressed format. Even if we could do that today, it is unlikely that it will still be the case in the future. By that time, access patterns, compression ratios, disk capacity, and other factors may require us to re-evaluate how the compression fits into our system anyway.

#### 4.2.2. Event Store redesign

The original design of the event store is very flexible and extensible, almost "too-flexible". The event store was designed and developed before the experiment started, and at that time it was not known how exactly it would be used. It has a number of features that are not used. All the extra features and powerful flexibility came with a price: space. With the redesigned event store, we estimated that size of event metadata could be reduced by as much as 80% (from 3.5 kB down to about 400 bytes) without losing any functionality, by redesigning persistent classes.

The event store was redesigned in the past few months. Size reduction was only one of the aspects that were addressed, others include:
- Introducing new features
- Making code and algorithms more robust
- Isolating persistent from transient world

We also found better ways to maintain all the flexibility that was offered by the old system. Further details on the redesigned are provided in [1].

### 4.3. Load balancing

The expansion of the analysis system inevitably introduced new issues that had to be dealt with. One of them was load balancing. In the past, it was relatively simple to balance load across few data servers. With many tens of servers it became a daunting task to arrange data in a way that evenly spread the load. The access patterns common in the HEP community are characterized by frequently changing hot-spots, making the task of load balancing very challenging.

An automatic load balancing system was developed to address this problem as a new extension to the AMS data server. The system is designed to be used for read-only data. It dynamically stages and/or replicates files based on configurable parameters, and host load. Each client first contacts a "*master*" AMS, which then redirects the client to the appropriate "*slave*" AMS. Slave AMSes can be taken offline transparently, which greatly increases fault tolerance of the whole system. Another important feature of automatic load balancing is its scalability, since it is hierarchical (it is easy to build a hierarchy: master of masters) - it is scalable.

Automatic load balancing is currently being tested and there are plans to put it in production in the next few months.

## 5. FUTURE OF BABAR DATA PERSISTENCY

A New BaBar Computing Model has been discussed in the past few months, aiming to address issues how to improve computing in BaBar. In particular it was examined how to improve data analysis and how to solve the growing problem of almost non-maintainable nTuple analysis. Some of the decisions made were directly related to data persistency.

In order to follow the current HEP trend as well as to allow interactive analysis in Root, a two-stage approach has been proposed to replace the existing Objectivity-bases event store. The "new-micro", based on Root I/O, is now believed to be a viable alternative to the nTuple analysis. The Condition Database, as well as Online Database would stay in the existing, Objectivity format.

The work on the new system has started. Existing code is reused as much as possible, and development progresses in light of experience with the existing system. It is expected that the new system will be ready for deployment in late 2003.

## 6. SUMMARY

The BaBar database system has managed to keep up with rapidly increasing B-Factory performance since it has been deployed in 1999. No major problems or showstoppers were found. We learned however, how challenging it is to keep up with growing size, complexity, and constantly changing demands.

Although the Objectivity-based event store has proven to be a good and working model for production activities, data analysis has remained a weak part of the system. Most users prefer to use much simpler nTuples, which they have been familiar with for years. This leads to producing unmaintainable number of copies of nTuples. In order to address this problem, a change in persistency format for the event store is planned.